\documentclass[sigconf,authorversion]{acmart}

\usepackage{xspace}
\usepackage[l2tabu,orthodox]{nag}
\usepackage[utf8]{inputenc}
\usepackage[british]{babel}
\usepackage{ifpdf}
\ifpdf
  \usepackage{microtype}
\fi

\usepackage{amsmath}
\usepackage[all]{onlyamsmath}

\makeatletter
\@ifclassloaded{acmart}{
}{
  \usepackage{newtxtext}
  \usepackage{newtxmath}
}
\makeatother

\usepackage[caption=false]{subfig}
\usepackage{booktabs}
\usepackage{upquote}
\usepackage{graphicx}
\usepackage{url}
\usepackage{bytefield}
\usepackage{listings}
\usepackage{algorithm}
\usepackage{algpseudocode}
\usepackage{color}
\usepackage{longtable}
\usepackage{hyphenat}

\usepackage{longtable}

\def\eg{\emph{e.g.,}\xspace}

\def\ie{\emph{i.e.,}\xspace}

\newcommand{\pb}[1]{\noindent{\bfseries \em #1}\hspace*{.3em}}

\newcommand{\key}[1]{\noindent{\emph{Key takeaway --- #1.}}\hspace*{.3em}}

\newcommand{\Web}{Web}

\newcommand{\EG}{e.g.,}
\newcommand{\ETLDpO}{eTLD$+$1}
\newcommand{\ttt}[1]{\path{#1}}
\newcommand{\RS}{Related Website Sets}

\newcommand{\NumStudyParticipants}{30}

\newcommand{\NumStudyErrors}{42} %
\newcommand{\PctStudyErrors}{36.8} %

\newcommand{\NumWrongStudyParticipants}{22}
\newcommand{\PctWrongStudyParticipants}{73.3}

\begin{document}

\title{A First Look at Related Website Sets}

\author{Stephen McQuistin}
\orcid{0000-0002-0616-2532}
\affiliation{
    \institution{University of St Andrews}
    \city{St Andrews}
    \country{UK}
}
\email{sm@smcquistin.uk}

\author{Peter Snyder}
\orcid{0000-0001-7880-2503}
\affiliation{
    \institution{Brave Software}
    \city{San Francisco}
    \country{USA}
}
\email{pes@brave.com}

\author{Hamed Haddadi}
\orcid{0000-0002-5895-8903}
\affiliation{
    \institution{Imperial College London \& Brave Software}
    \city{London}
    \country{UK}
}
\email{h.haddadi@imperial.ac.uk}

\author{Gareth Tyson}
\orcid{0000-0003-3010-791X}
\affiliation{
    \institution{Hong Kong University of Science \& Technology (GZ)}
    \city{Guangzhou}
    \country{China}
}
\email{gtyson@ust.hk}

\copyrightyear{2024}
\acmYear{2024}
\setcopyright{acmlicensed}
\acmConference[IMC '24]{Proceedings of the 2024 ACM Internet Measurement Conference}{November 4--6, 2024}{Madrid, Spain}
\acmBooktitle{Proceedings of the 2024 ACM Internet Measurement Conference (IMC '24), November 4--6, 2024, Madrid, Spain}
\acmDOI{10.1145/3646547.3689026}
\acmISBN{979-8-4007-0592-2/24/11}

\settopmatter{printacmref=true}



\begin{abstract}
We present the first measurement of the user-effect
and privacy impact of "Related Website Sets," a recent
proposal to reduce browser privacy protections
between two sites if those sites are related to each other.
An assumption (both explicitly and implicitly) underpinning the
Related Website Sets proposal is that users can accurately determine
if two sites are related via the same entity. 
In this work, we probe this assumption via measurements and a user study of
\NumStudyParticipants{} participants, to assess the ability of Web
users to determine if two sites are (according to the \RS{} feature)
related to each other. We find that this is largely
not the case. Our findings indicate that \NumStudyErrors{} (\PctStudyErrors{}\%)
of the user determinations in our study are incorrect in privacy-harming ways, where users think that sites are not related, but would be treated as related (and so
due less privacy protections) by the \RS{} feature. Additionally,
\NumWrongStudyParticipants{} (\PctWrongStudyParticipants{}\%) of participants
made at least one incorrect evaluation during the study.
We also characterise the \RS{} list, its composition over time, and its governance.

\end{abstract}

\maketitle

\section{Introduction}
\label{sec:introduction}

Browser vendors increasingly implement site partitioning (sometimes called third-party storage partitioning) into their products to protect user privacy on the \Web{}. While browser vendors seem to agree that that partitioned third-party state should
be the default in all browsers, they largely \emph{disagree} on what steps, if any, should be taken in the interim, while websites adjust.
Many websites today were designed to run without storage partitioning,
and so break (in the subjective evaluation of the site operator, the site user, or both) when storage partitioning is applied. Some browsers have decided to prioritise user privacy, while others apply heuristic and list-based approaches to determine when reduced privacy protections are acceptable.

Google has proposed one such list-based approach for deciding when to (or, not t to) apply storage partitioning, called \RS{}. The \RS{} proposal consists of two parts. First,
a list of sites that are related to one another, and second, a browser policy for allowing unpartitioned storage
access between sites that the list indicates are related to each other. 

Underlying the \RS{} proposal is the intuition that if users understand that two sites are
affiliated to a common organisation, then there is less need for the browser to enforce a privacy
boundary between those two sites. Under this view, enforcing a privacy boundary between two sites that a user knows are related to each other is harmful to users
(\eg risk of sites breaking, needing to redundantly log into multiple related sites),
without providing any privacy improvement; the user expected that their information was going
to be shared by both sites anyway. Crucially, websites related to each other under the \RS{} proposal do not need to have a common owner, enabling data sharing that would otherwise not occur.

In this work, we evaluate whether the assumptions underlying the \RS{} proposal are
accurate. Understanding whether users perceive site relationships in the same way
that \RS{} list maintainers do is important for Web privacy. If user perceptions
\emph{do not} match the list maintainers' expectations, \RS{} will result in
privacy- (and user-)harming behavior, just as the Web is seemingly about to adopt a new
privacy-improving baseline.

We make the following contributions:

\begin{enumerate}
    \item A \textbf{user study} where participants are asked to subjectively evaluate whether
        pairs of sites are related to each other (\S\ref{sec:relatedness-survey});%
    \item An \textbf{evaluation of the composition and management of the current \RS{} list} (\S\ref{sec:set-submissions}); and
    \item A discussion of \textbf{how \RS{} relates to other proposals}, from both
        other browser vendors, and other privacy tools (\S\ref{sec:related-work}).
\end{enumerate}

\pb{Reproducibility and data access.} We make available our code for gathering, processing, and analysing the data discussed in this paper. This, and the data used in our survey, along with its anonymised results, is available from \url{https://doi.org/10.17630/450a41e5-1f12-43ef-b909-2640dcd0fe50}.

\section{Web privacy boundaries}
\label{sec:background}

\pb{Site-as-privacy-boundary on the Web.} All current Web browsers either use, or plan to use, the \emph{site} as the default
privacy boundary on the Web. The site, in this context, refers to ``effective top level
domain, plus one subdomain'' (\ETLDpO{}). The ``effective top level domain''
refers to the suffixes defined in the public suffix
list.\footnote{\url{https://publicsuffix.org/}}

Web browsers use \ETLDpO{} domains as site boundaries and aim to keep
activity on one site unlinkable to activity on any other site. For example,
the browser aims to keep activity on \texttt{facebook.com} unlinkable from activity on
\texttt{mayoclinic.com} (two different sites), but does not aim to keep activity on
\texttt{eff.org} unlinkable to activity on \texttt{act.eff.org} (two domains on the same site).

\pb{Enforcing site-as-privacy-boundary.} Storage partitioning is a general strategy of giving sites access to different storage areas (\EG{} cookies, \texttt{localStorage}) depending on the context the site is loaded in, and is the primary way that browsers enforce (or plan to enforce) the site-as-privacy-boundary.

As an example, imagine that \ttt{tracker.example} is a domain operated by a tracking service. This site can be loaded in different contexts: for example, users can visit it directly, as the first party, or they can access it indirectly, via assets (such as adverts) loaded in an \ttt{<iframe>}. Without storage partitioning, \ttt{tracker.example} is able to set and access the same set of cookies in both of these contexts, allowing it to track users across the Web.

Storage partitioning prevents this scenario, and enforces the site's privacy boundary
by giving \ttt{tracker.example}~\cite{jueckstock2022measuring} access to a different set of cookies depending on the context it is loaded in. When the user visits
\ttt{tracker.example} directly, it can access one set of cookies, and when \ttt{tracker.example} is being loaded as a third-party on another site, such as \ttt{site.example}, it can access a separate set of cookies.

\pb{\textit{Related Website Sets} and creating exceptions to the site-as-privacy-boundary.}Google's proposed \emph{Related Website Sets} feature is a browser capability that creates exceptions to storage partitioning, weakening the site-as-privacy-boundary policy. The general idea behind the Related Website Sets proposal is that there is little benefit to users (and possibly, some inconvenience created) by enforcing a privacy boundary between different sites that are clearly affiliated to the same organisation.

To understand how the Related Website Sets would work in practice, consider the following real-world example.
``Times Internet'' is a company that operates multiple popular websites in India, including
\ttt{https://timesinternet.in} and \ttt{https://www.indiatimes.com/}. Without Related Website Sets, the browser would not allow either site to know that the same user was interacting with both sites until
the user took some additional step, such as logging into both sites with common user credentials.
With Related Website Sets, either site can embed an iframe from the other site. If code in that iframe then calls the \ttt{requestStorageAccess} method, then, since the two sites are related by the Related Website Sets list,\footnote{\url{https://github.com/GoogleChrome/related-website-sets/blob/main/related_website_sets.JSON}} the browser would allow code running in that iframe access to that site's unpartitioned storage area (\eg cookies, user identifiers, etc),
despite being embedded as a third-party. This would allow
both sites to link page visits on each site to the same user.

Related Website Sets are comprised of multiple subsets. \emph{Service sites}, which cannot be the top-level domain in a storage access grant. Users must first interact with another member of the set; otherwise, storage access is automatically granted. Service sites must be under common ownership with the set primary, and are designed to support the functionality or security of other set members. 
\emph{Associated sites} are sites that must be clearly affiliated with the set primary (\eg using common branding, an about page, or similar). However, they are \emph{not} required to have common ownership. \emph{ccTLD sites} are ccTLD variations of other set members, and must have common ownership with the domain that they are a variant of.

\pb{Status of site-as-privacy-boundary in browsers.} 
Most browsers currently use the site as the Web's privacy boundary, or have plans to. Safari, Brave, and Firefox all enforce the site-as-privacy-boundary by default, though some 
make some temporary exceptions to avoid breaking websites.
Firefox and Safari both include the Storage Access 
API\footnote{\url{https://privacycg.github.io/storage-access/}} that allows embedded third-party sites to request unpartitioned storage (and so, for an exception to the site-as-privacy-boundary 
policy), but requires user consent via prompt in some (Firefox) or all (Safari) cases.

Chrome and Edge \emph{do not} currently implement a default site-as-privacy-boundary policy. While Chrome will continue to allow third-party cookies for the immediate future, it has deployed Related Website Sets, and intends it to be a permanent method that sites can use to gain exceptions from the site-as-privacy-boundary policy where users have opted to enable it\footnote{\url{https://privacysandbox.com/intl/en_us/news/privacy-sandbox-update/}}. To the best of our knowledge, Microsoft has not yet announced if it plans to adopt Related Website Sets.

\pb{Governance of Related Website Sets.} Related Website Sets are proposed by site owners using pull requests on GitHub\footnote{\url{https://github.com/GoogleChrome/related-website-sets}}. Pull requests are subject to a series of automated and manual checks. 
The automated checks ensure that Google's Contributor Licence Agreement\footnote{\url{https://opensource.google/documentation/reference/cla/}} has been completed by the contributor, before a series of technical validation checks are run (\eg ensuring that no non-HTTPS sites are present, that all sites are \ETLDpO{}s, among other factors\footnote{\url{https://github.com/GoogleChrome/related-website-sets/blob/main/RWS-Submission_Guidelines.md\#set-validation-requirements}}).

\section{Can Users Determine Relatedness?}
\label{sec:relatedness-survey}

Underpinning the Related Website Sets proposal is the motivation that it creates exceptions to the site-as-privacy-boundary where doing so removes inconvenience to users, and follows from their expectations of privacy. As described in Section~\ref{sec:background}, the proposal allows sites to be related to each other by common ownership \emph{or} common affiliation. The associated subset is reserved for ``\emph{domains whose affiliation with the set primary is clearly presented to users}''\footnote{\url{https://github.com/GoogleChrome/related-website-sets/blob/main/RWS-Submission_Guidelines.md##set-formation-requirements}}. Therefore, the efficacy of the privacy boundaries that the Related Website Sets approach constructs relies upon users being able to determine that set members are related to each other, regardless of common ownership, and therefore, that they could reasonably expect their browser to share data between them.

\pb{Can users accurately determine relatedness?} 
To assess whether users could determine that websites were related to each other, we conducted a user study in May 2024. Users were presented with links 20 pairs of websites, asked to open those links and to view the websites, and then asked to determine if the two websites were related to each other by an affiliation to a common company or organisation. Users were not asked to identify what, if any, the common affiliation was. Each question was timed to determine how long participants spent assessing the relatedness of each pair of sites. Finally, after answering all 20 questions, participants were asked to indicate which factors they considered in determining when websites were and were not related to each other. Participants were able to exit the survey at any time, and to skip individual questions. The survey was entirely anonymous, with no personally identifiable information collected about participants, ethical approval was obtained,\footnote{Approved by the University Teaching and Research Ethics Committee (UTREC) at the University of St Andrews, with approval code CS17715.} and best practice related to informed consent was followed in carrying out the study.

The study was advertised via social media and within the institutions of the authors. This may skew participation towards individuals that have a computer science background, and that a familiar with the Web. While we do not investigate the impact that this has on our results, we hypothesise that they represent a baseline, and that participants with less familiarity would be less able to determine the relatedness of websites.   

The pairs of websites were drawn from 4 groups, with each participant asked about 5 pairs, at random, from each:
\begin{enumerate}
    \item \emph{Sites that are members of the same Related Website Set}. All combinations of set primaries and associated sites within each set (``RWS (same set)''). This combinations in this group \emph{are} related under the RWS proposal.
    
    \item \emph{Sites that are members of other Related Website Sets}. All combinations of set primaries and associated sites; each site from a different set (``RWS (other set)''). The combinations in this group \emph{are not} related under the RWS proposal.
    
    \item \emph{Sites from Related Website Sets and another site within the same Forcepoint category}. Pairs were formed from all combinations of set primaries and associated sites, and a list of 200 sites, drawn randomly from the Tranco Top 10K list~\cite{pochat2018tranco}, filtered to sites within the same Forcepoint\footnote{The Forcepoint ThreatSeeker (\url{https://www.forcepoint.com/product/feature/threatseeker}) database classifies URLs into broad categories (\eg news and media, business and economy).} category (``Top Site (same category)''). The combinations in this group are not related under the RWS proposal, but may be similar to each other given that they fall within the same Forcepoint category.
    
    \item \emph{Sites from Related Website Sets and another site in a different Forcepoint category}. All combinations of set primaries and associated sites, and the above list of 200 sites, filtered to sites in a different Forcepoint category (``Top Site (other category)''). The combinations in this group are not related under the RWS proposal, but may be dissimilar to each other given that they are in different Forcepoint categories.
\end{enumerate}

\begin{table}
\begin{tabular}{l|rrrrrr}
\toprule
\textbf{Category} & \textbf{Related} & \textbf{Unrelated} \\
\midrule
RWS (same set) & 72 (28.1s) & 42 (39.4s) \\
RWS (other set) & 5 (25.5s) & 100 (32.5s) \\
Top Site (same category) & 8 (32.6s) & 104 (33.2s) \\
Top Site (other category) & 7 (31.5s) & 92 (26.5s) \\
\bottomrule
\end{tabular}

\caption{Website relatedness survey results summary.}
\label{tab:survey-summary}
\end{table}

\begin{figure*}
  \begin{minipage}{.3\textwidth}
        \centering
        \includegraphics[width=0.8\textwidth]{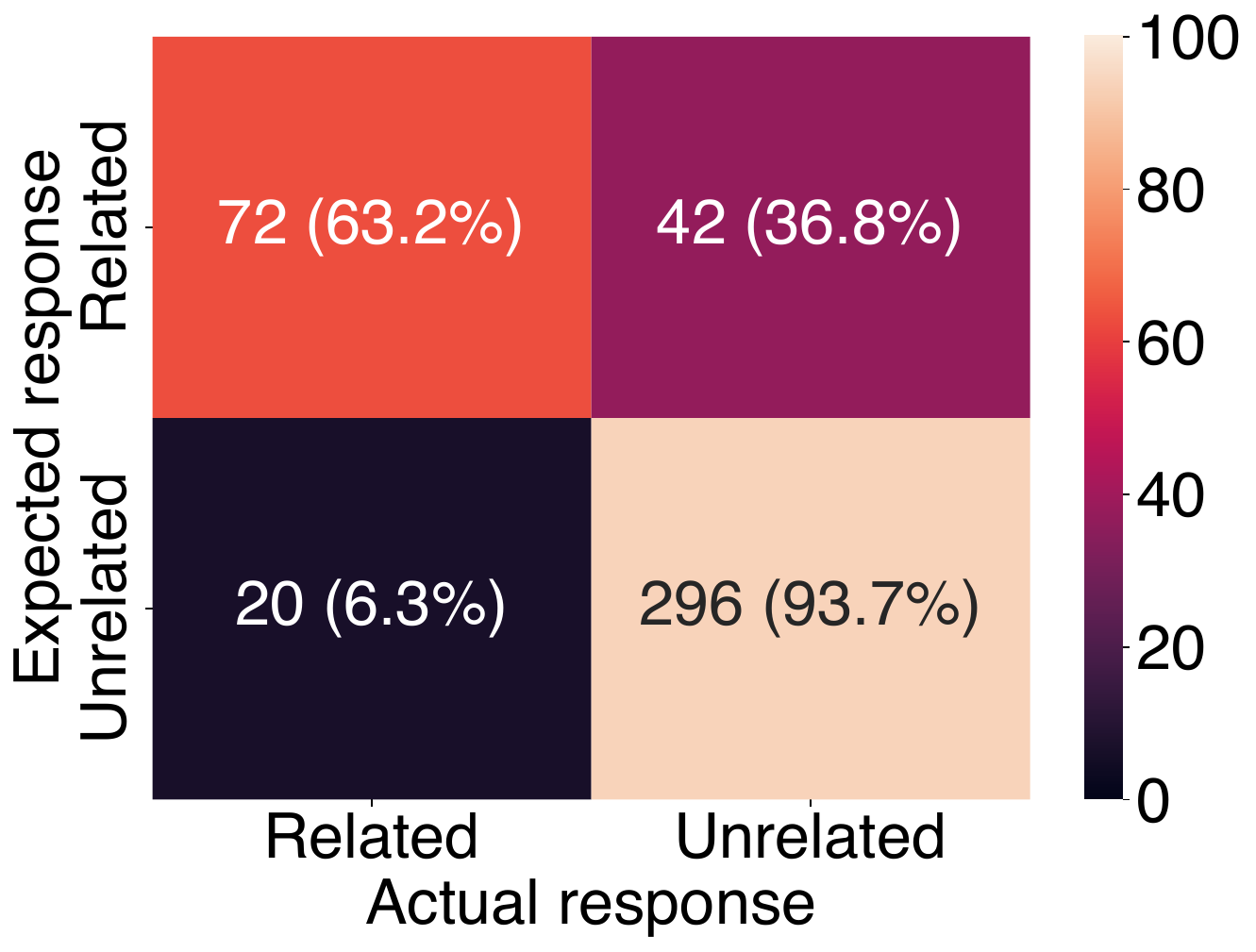}
        \caption{Website relatedness survey results matrix; percentages and heatmap color are within \emph{Expected response}.}
        \label{fig:survey-matrix}
  \end{minipage} \quad
  \begin{minipage}{.3\textwidth}
        \centering
        \includegraphics[width=\textwidth]{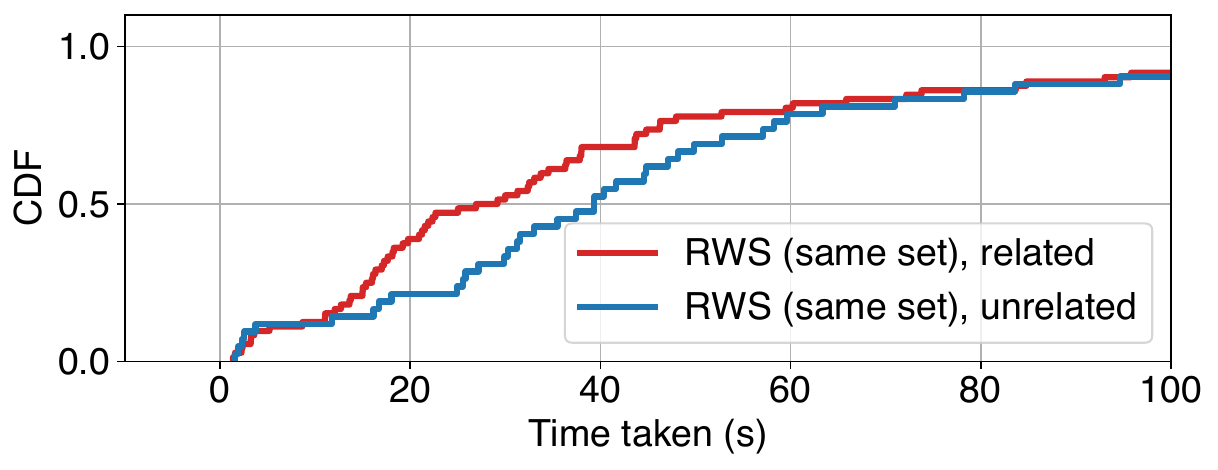}
        \caption{Website relatedness survey timing distributions; for pairs within the \emph{RWS (same set) category}, split by response.}
        \label{fig:survey-time-cdf-rws_sameset}
  \end{minipage} \quad
  \begin{minipage}{.3\textwidth}
        \centering
        \vspace{3mm}
        \includegraphics[width=\textwidth]{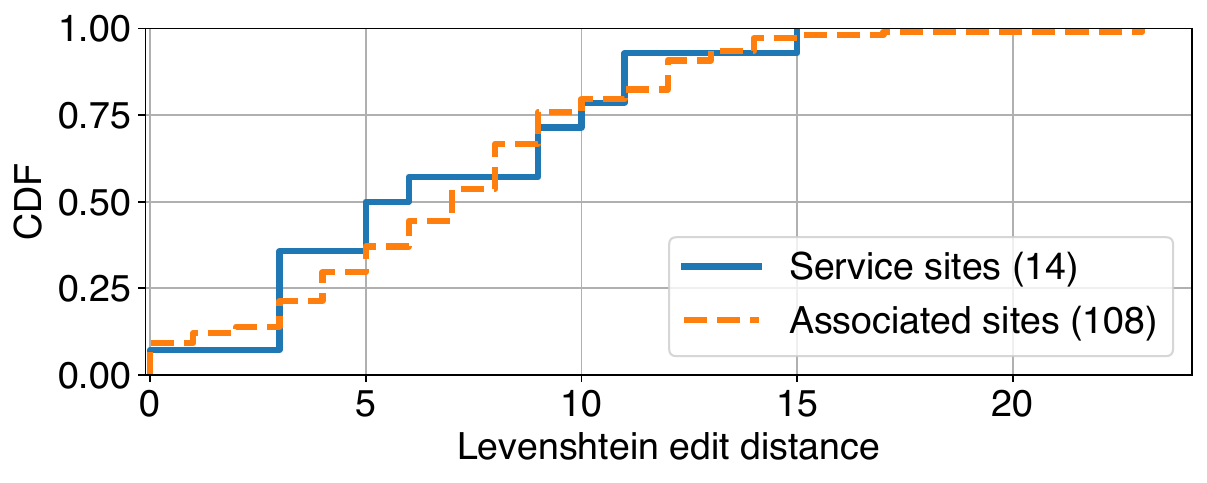}
        \caption{CDFs of the Levenshtein edit distance between service/associated sites and their primary domain, per the Related Website Set list at 26 March 2024.}
        \label{fig:set-levenshtein-latest}
  \end{minipage}
\end{figure*}

\begin{table}
\begin{tabular}{lrr}
\toprule
\textbf{Factor used} & \textbf{Related} & \textbf{Unrelated} \\
\midrule
\small
Domain name & 12 (57.1\%) & 11 (52.4\%) \\
Branding elements & 14 (66.7\%) & 13 (61.9\%) \\
Header text & 9 (42.8\%) & 11 (52.4\%) \\
Footer text & 13 (61.9\%) & 11 (52.4\%) \\
``About'' pages or similar & 10 (47.6\%) & 7 (33.3\%) \\
Other & 4 (19\%) & 5 (23.8\%) \\
\bottomrule
\end{tabular}

\caption{Website relatedness survey: factors used to determine relatedness and unrelatedness.}
\label{tab:survey-relatedness}
\end{table}

Further manual filtering was performed to check that the website on the Related Website Sets list were live, and that they were primarily English-language. As the survey was advertised in English-speaking regions, this manual filtering was performed to ensure that participants could reasonably assess relatedness. A large proportion of the sites in the Related Website Sets are not primarily English-language, and so this filtering reduced the number of sites on the Related Website Sets list from 146 sites to 31 sites. 822 pairs were generated, comprised of 39 \emph{RWS (same set)} pairs; 426 \emph{RWS (other set)} pairs; 141 \emph{Top Site (same category)} pairs; and 216 \emph{Top Site (other category)} pairs. The full set of generated pairs, alongside the anonymised data gathered by the study, is contained in the dataset released alongside this paper, and described in Section~\ref{sec:introduction}.

A total of 30 participants\footnote{Due to the anonymous nature of the survey, ``participants'' here means individual sessions of the survey.} provided 430 responses. Figure~\ref{fig:survey-matrix} and Table~\ref{tab:survey-summary} summarize the results. Of the 114 responses to pairs within the same RWS set (\ie those pairs that are related), 36.8\% incorrectly identified the websites as being unrelated. Across the 316 responses for pairs drawn from the other 3 categories, 93.7\% indicate that the websites are unrelated. 

Performing a two-sample Kolmogorov-Smirnov test pair-wise across the timing distributions for responses within each of the categories, we find no statistical significance between them ($p < 0.05$). However, looking only at the split of responses to pairs within the \emph{RWS (same set)} category, as shown in Figure~\ref{fig:survey-time-cdf-rws_sameset}, we find a statistically significant difference in the time taken to determine relatedness vs. unrelatedness. This suggests that participants were more quickly able to determine relatedness, and that they spent longer evaluating the websites before concluding that they were unrelated.

\key{In 36.8\% of pairs, participants incorrectly identified that websites drawn from the same Related Website Set were unrelated, and, when presented with such websites, spent longer determining their relatedness. Users would be unlikely to expect data to be shared in those instances where they were unable to determine relatedness}

\pb{How do users determine relatedness?} Table~\ref{tab:survey-relatedness} summarizes the factors that participants used to determine website relatedness. Of the 21 participants that responded to this question, ``Branding elements'' (\eg logos, colors, and similar) were most frequently used. The domain name itself was also frequently used, with 57.1\% of respondents using this to determine that websites were related. 

\key{Common branding elements, along with the domain names of the sites, are often used by participants to determine website relatedness. These factors should be used by the maintainers of the Related Website Set list when determining whether associated sites should be included in a set}

\pb{How similar are the second-level domains of set members?} \\
Given that participants (57.1\% of respondents) indicate that they assessed relatedness using the domain names of sites, we assess whether automated tooling could be used to determine relatedness. To estimate how feasible this is, Figure~\ref{fig:set-levenshtein-latest} shows CDFs of the Levenshtein edit distance between each service or associated site's SLD, and its set primary's SLD. 

A small proportion (9.3\%) of associated site SLDs are identical to that of their set primary (\eg \texttt{poalim.xyz} is associated with \texttt{poalim.site}). 
This is likely to be impacted by the ability for site owners to declare an unlimited number of ccTLD variants of domains within the same set, limiting exact-match SLDs to those that have different gTLDs. 
Associated site SLDs have a median edit distance of 7 from that of their set primary's SLD. 
In such cases, it is unlikely a user could easily identify that the sites as related.
Within this, there are sites that share common components (\eg \texttt{autobild.de} associated to \texttt{bild.de}) and others that are entirely distinct (\eg \texttt{nourishingpursuits.com} associated to \texttt{cafemedia.com}). In addition, though not present in the Related Website Sets list, domain squatting means that using SLD similarity as a measure of relatedness is risky; it does not confirm common ownership in itself.

\key{The similarity of SLDs is not a reliable way of determining relatedness between an associate site and a set primary, with half of associated site SLDs having an edit distance of 6 or more from that of their set primary}

\begin{figure*}
  \begin{minipage}{.3\textwidth}
        \centering
\includegraphics[width=\textwidth]{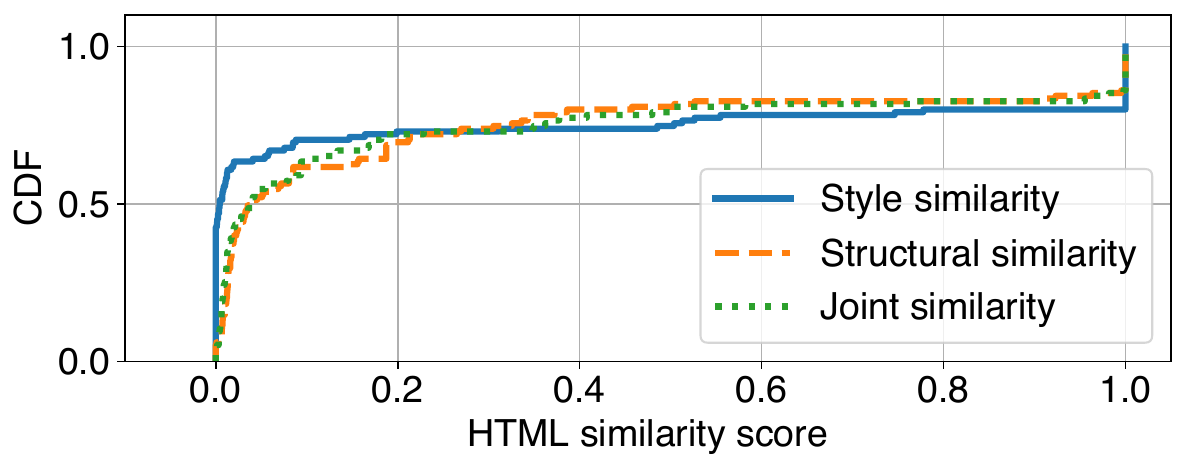}
\caption{CDFs of HTML similarity scores of set primaries and their service/associated sites, per the RWS list at 26 March 2024.}
\label{fig:set-html-sim-latest}

  \end{minipage} \quad
  \begin{minipage}{.3\textwidth}
        \centering
\includegraphics[width=\textwidth]{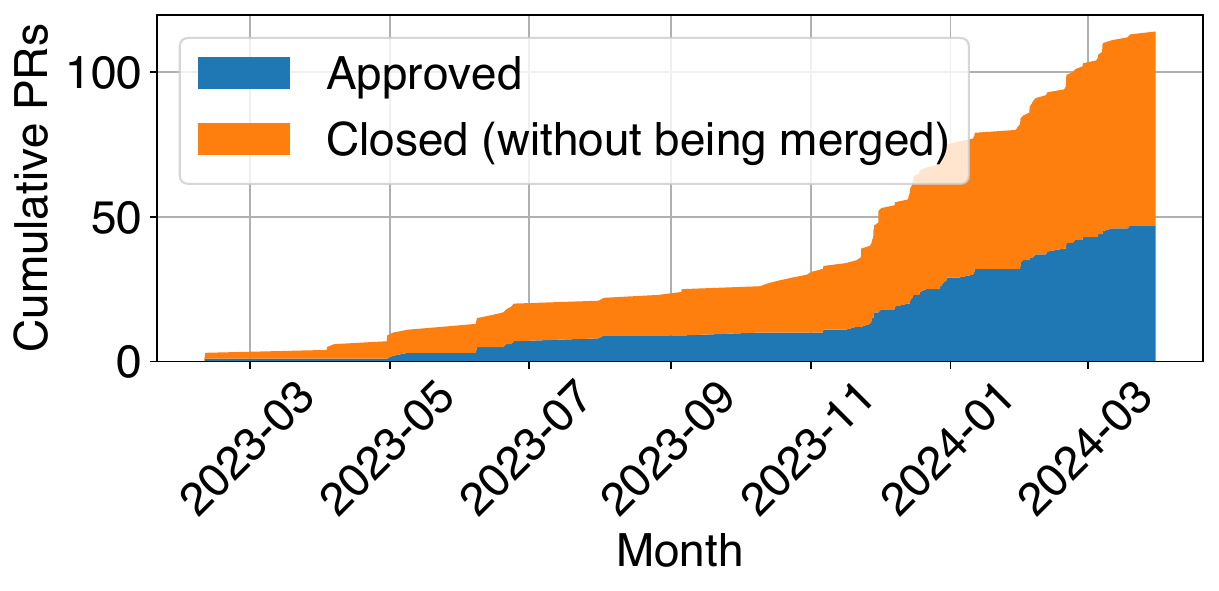}
\caption{Cumulative count of PRs that propose a new set, split by final state.}
\label{fig:pr-approvals-cumulative}
\end{minipage} \quad
  \begin{minipage}{.3\textwidth}
        \centering
\includegraphics[width=\textwidth]{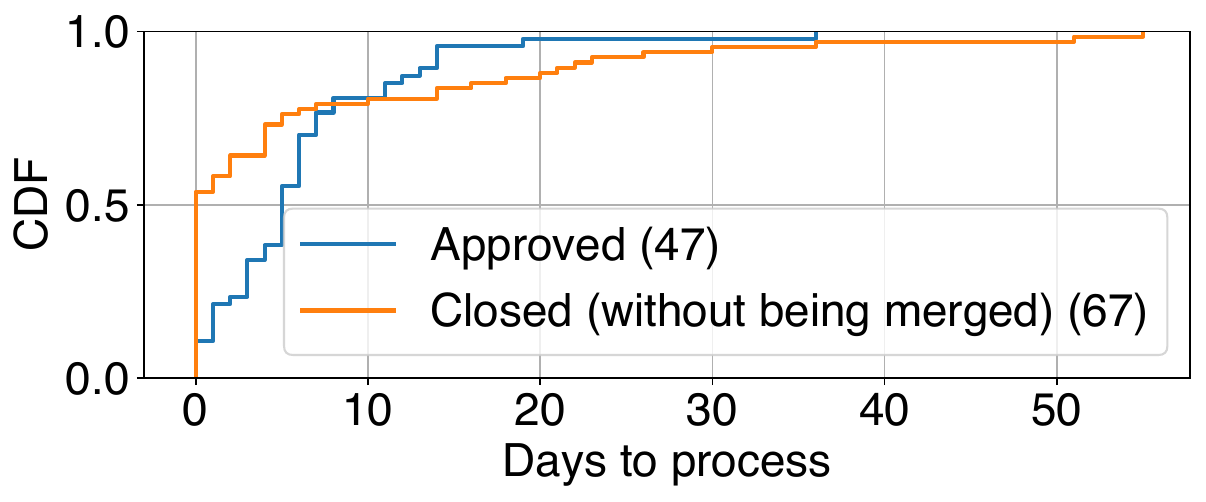}
\caption{CDF of days taken to process PRs that propose a new set.}
\label{fig:pr-approvals-days_to_proc}
  \end{minipage}
\end{figure*}

\pb{How similar in structure and style are set members?} 
Next, with 66.7\% of respondents indicating that they used common branding elements to determine relatedness, we assess whether the \emph{content} of the sites is similar, by computing the HTML similarity of each service and associated site when compared to its set primary.
We use a well-known library\footnote{\url{https://github.com/matiskay/html-similarity}} that, for a pair of websites, can compute the \emph{style similarity} (based on CSS classes), \emph{structural similarity} (based on HTML tags), and \emph{joint similarity} (a weighted sum of both).
From Figure~\ref{fig:set-html-sim-latest} we observe that a significant proportion of service
and associated sites
are \emph{dissimilar} to their set primaries, with a median joint HTML similarity score of 0.04.

\key{HTML similarity metrics show that service and associated sites are largely dissimilar to their set primaries. This means that manual validation of the common affiliation of associated sites is necessary, given that it is difficult to assess automatically}

\section{The Related Website Set List}
\label{sec:set-submissions}

Having explored the question of the extent to which users can determine website relatedness, we next characterise the current Related Website Sets list, and how it is managed using GitHub. These are important research questions: if widely adopted, the composition and management of the Related Website Sets list may have significant implications on user privacy.

\begin{figure*}
  \begin{minipage}{.3\textwidth}
        \centering
\includegraphics[width=\textwidth]{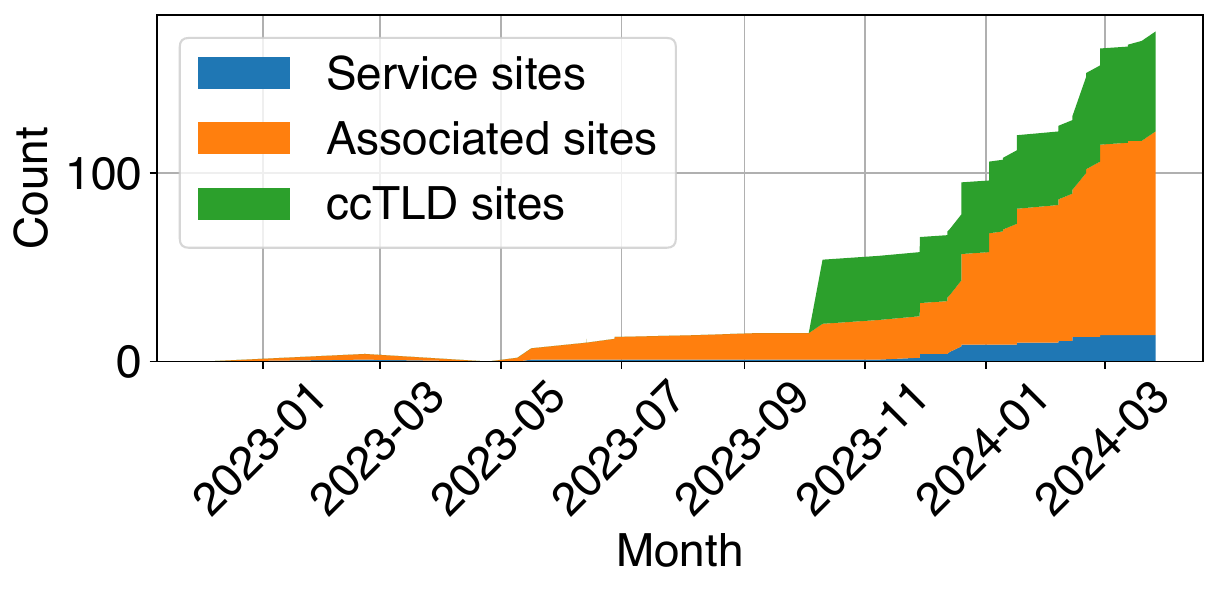}
\caption{Set composition over time.}
\label{fig:set-composition-latest}
  \end{minipage} \quad
  \begin{minipage}{.3\textwidth}
        \centering
\includegraphics[width=\textwidth]{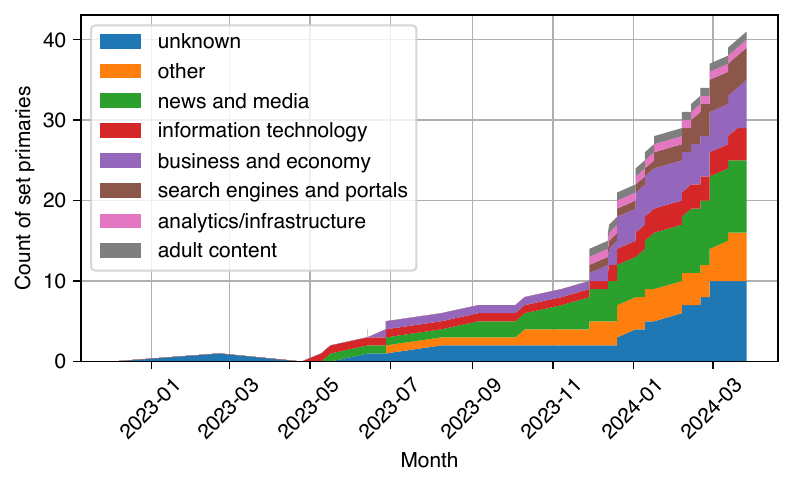}
\caption{Forcepoint ThreatSeeker categories of set primaries}
\label{fig:set-cats-primary}
\end{minipage} \quad
  \begin{minipage}{.3\textwidth}
        \centering
\includegraphics[width=\textwidth]{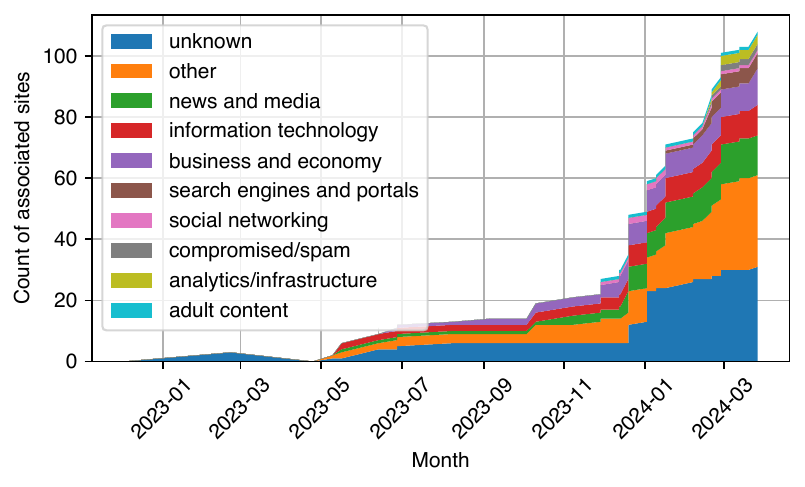}
\caption{Forcepoint ThreatSeeker categories of associated sites.}
\label{fig:set-cats-associated}
  \end{minipage}
\end{figure*}

\pb{How are set members distributed across subsets?} We first consider the composition of sets in terms of the subset types defined in Section~\ref{sec:background}. Associated sites are the most potentially privacy-impacting, given that common ownership is not required, and that users often fail to determine relatedness.
Figure~\ref{fig:set-composition-latest} shows the count of sites per subset category. As of the most recent RWS list in the dataset (26 March 2024), there were 41 sets; of these, 22\% had one or more service sites; 14.6\% had one or more ccTLD sites; and 92.7\% had one or more associated sites. This shows that the overwhelming use case for the Related Website Sets mechanism is to incorporate associated sites, with a mean of 2.6 associated sites per set.

\key{92.7\% of Related Website Sets include one or more associated sites, where common ownership to the set primary is not required}

\pb{What categories of sites are sets made up of?} Next, we explore the composition of sets in terms of the Forcepoint category that they fall into, to characterise those sites that are seeking an alternative to third-party cookie functionality. Figures~\ref{fig:set-cats-primary} and \ref{fig:set-cats-associated} show the categories of set primaries and associated sites, drawn from Forcepoint ThreatSeeker classifications.
Note, similar categories are merged together, while smaller categories are grouped into ``Other''. As shown, the largest individual category for set primaries is \emph{News and media}; these contain associated sites in other categories (\eg the set for \texttt{bild.de}, a German news site, includes \texttt{computerbild.de}, its related IT news website). The RWS mechanism allows data to be shared across these sites, enabling, for example, common ad tracking and profiling. Additionally, some sets contain analytics or tracking infrastructure explicitly: \texttt{ya.ru} (Yandex, a Russian Internet company) includes \texttt{webvisor.com}, a Web analytics service.

\key{A significant portion of sites fall into categories, such as ``News and media'', that are likely to benefit from existing third-party cookie functionality, making their early adoption of alternatives intuitive}

\pb{How often are set proposals rejected?} Next, we turn to the governance of the Related Website Sets list. Given the potential privacy implications of the list, it is crucial that it is well maintained, and that the rules governing set submissions are clear. Figure~\ref{fig:pr-approvals-cumulative} shows the cumulative count of pull requests on the Related Website Sets list over time, through to March 30th 2024, comprising 114 requests. As shown, the rate at which pull requests are submitted has grown over time, as the Related Website Sets proposal has developed. Additionally, the split between approved and closed (without merging) requests has shifted, with 58.8\% of all pull requests closed without being merged, suggesting a significant volume of invalid submissions.

A set may be proposed across multiple pull requests. Site owners will often propose a new set, receive the results of the automated checks, close the pull request, and then open a new pull request. Across the 114 pull requests in the dataset, only 60 set primaries are represented, giving a mean of 1.9 pull requests per set primary.

\key{58.8\% of pull requests on the Related Website Sets list are rejected, suggesting that the automated checks are successful in enforcing the technical set-level requirements of the Related Website Sets proposal}

\begin{table}
\begin{tabular}{lr}
\toprule
\textbf{GitHub bot comment} & \textbf{Count} \\
\midrule
\small
Unable to fetch .well-known JSON file & 202 \\
Associated site isn't an eTLD+1 & 65 \\
Service site without X-Robots-Tag header & 19 \\
PR set does not match .well-known JSON file & 12 \\
Alias site isn't an eTLD+1 & 10 \\
Primary site isn't an eTLD+1 & 9 \\
Other & 8 \\
No rationale for one or more set members & 5 \\
\bottomrule
\end{tabular}

\caption{RWS GitHub bot validation messages.}
\label{tab:pr-validation-messages}
\end{table}

\pb{What are the common validation errors?} The set-level technical validation checks are carried out automatically, and the results are reported back via a GitHub bot, which adds a comment to the pull request; understanding the reasons why set proposals are rejected may help to streamline the process, and identify common misunderstandings. Table~\ref{tab:pr-validation-messages} shows the number of occurrences of each of the validation errors observed in the pull request dataset. While validation is performed at the set-level, some error messages are produced per site; additionally, the validation is performed again if the pull request is updated. This results in a one-to-many mapping between pull requests and the validation errors that are observed.

The most frequent error is that the \texttt{.well-known} file is not able to be fetched. This is a JSON file (to be publicly accessible via each set member) that contains the set (i.e., the same data available in the Related Website Set list).
This ensures that proposers have administrative access to the domains that they are submitting. This is likely to be an oversight on the part of the submitter.

The next most frequent error is that the set contains an associated site that is not an eTLD+1. For example, a set proposer might have \texttt{example.com} as the set primary, and \texttt{a.example.com} as an associated site. Assuming that \texttt{example.com} is not an eTLD (\ie present in the Public Suffix List), then \texttt{a.example.com} is not a third-party site, with respect to \texttt{example.com}. In these cases, this represents a fundamental misunderstanding of the privacy boundaries that already exist, and that Related Website Sets are reshaping.

\key{The most frequent validation errors suggest that the Related Website Sets proposal is complex, both in terms of the technical requirements (\eg for the \texttt{.well-known} file), and the privacy boundaries that are constructed. This suggests that documentation and tooling (for validating a proposed set before submission) could be improved}

\pb{How long does it take for a pull request to be processed?} \\ 
Having looked at the rate that PRs are submitted, and the common automated validation errors, Figure~\ref{fig:pr-approvals-days_to_proc} shows the time taken for set proposals to be processed, either successfully (\ie approved and merged in) or unsuccessfully (\ie closed, without being merged in).

We see that 54.3\% of unsuccessful pull requests are closed within the day that they are opened. This is likely to result from the automated validation process described above: submitters frequently close their pull requests after receiving output from the GitHub bot.
However, we observe a long-tail in the time taken to close unsuccessful requests.

The median time to process a successful request is 5 days. Only 1 of the 47 merged pull requests fail any of the automated checks, suggesting that the time to process successful requests is driven primarily by their manual validation by the maintainers of the Related Website Sets list.

\key{The automated validation checks provide quick feedback to submitters, while the manual validation checks contribute to a median time to process successful requests of 5 days. Given the low rate of requests, it is unclear how the manual component of the process will scale should this mechanism become more widely adopted}

\section{Related Work \& Discussion}
\label{sec:related-work}

\pb{List-based Web privacy.} \RS{} is most similar to the Disconnect
list\footnote{\url{https://github.com/disconnectme/disconnect-tracking-protection}},
an expert-curated commercial product. The Disconnect
list consists of two sub-lists: the \emph{services} list,
a list of domains determined to be related to privacy-harming
(or otherwise undesirable), and the \emph{entities} list, a list of domains that are run by the same organizations.
This second list is similar
to Google's \RS{} in several ways. First, both list sets of domains that are controlled by the same organization. Second, both are used by popular Web browsers (\eg Firefox and Edge) to
decide whether privacy protections should be relaxed.
Third, both lists are curated by a small group of experts. 
Google employees. 
A crucial difference is that the \RS{} list, through associated sites, relaxes the requirement that sites be operated by the same company, and only requires that they have a common affiliation that is clearly presented to users. Our work has shown that this relaxation is often at odds with users' ability to determine website relatedness.

There are other popular list-based approaches to Web privacy
that differ from the \RS{} list significantly. Filter lists, such as EasyList\footnote{\url{https://easylist.to/}}, and supplementary 
lists like uBlock
Origin\footnote{\url{https://github.com/gorhill/uBlock}} and 
AdGuard\footnote{\url{https://adguard.com/en/welcome.html}}, are primarily
crowd-sourced. While the \RS{} list
defines rules over domains, filter lists define rules over URLs.
Filter lists are much larger than expert curated lists; \RS{} and 
Disconnect describe hundreds or thousands of domains (respectively), while
EasyList alone includes tens-of-thousands of rules~\cite{snyder2020filters}.

\pb{Stateful third-party Web tracking.} The \RS{} proposal, and this work examining it, relates to the
well-studied area of stateful third-party tracking on the Web.
Browser state could be abused to violate user privacy on the Web,
both through browser capabilities intended to store application level
information (\ie cookies~\cite{mayer2012third, 
roesner2012detecting}, and user IDs), but also other indirect capabilities (\eg
browser caches~\cite{jackson2006protecting}, Flash~\cite{soltani2010flash}, cached \ttt{ETag} headers~\cite{ayenson2011flash}, the DNS cache~\cite{klein2019dns},
the ``favicon'' cache~\cite{solomos2021tales}, ``Alternative-Service'' 
(\ttt{Alt-SVC}) headers~\cite{tiwari2019alternative}, and ``HTTP Strict 
Transport Security'' (\ttt{HSTS}) headers~\cite{syverson2018hsts}).
Other work attempted to distinguish
between cases where third-party state is used to track users, and cases where third-party state is used for more benign purposes~\cite{li2015trackadvisor}. 
Our work has shown that the RWS may open up users to greater tracking.

\section{Conclusions}
\label{sec:conclusions}
This paper has evaluated whether the assumptions underlying the RWS proposal are accurate.
Browser vendors have enabled, or plan to enable, third-party storage partitioning, an effective protection against the most pervasive kinds of privacy harm on the Web. However, there are significant compatibility risks. Many sites on the Web were designed for how Web browsers worked when they were less privacy-protected, and efforts to find ways to improve privacy \emph{without} breaking ``legacy'' sites are essential.

While \RS{} proposes an appealing solution to this problem, we find that its underlying assumptions about \emph{relatedness} do not hold, and we find that Web users cannot accurately evaluate whether two sites are in fact operated by the same organisation. This suggests that exceptions made to the site-as-privacy-boundary, on the basis of relatedness, need to be explicitly indicated to the user (\eg via the browser UI itself); we leave study of the efficacy of such an approach to future work.

We hope these findings are useful to those developing techniques to improve Web privacy without breaking compatibility, and demonstrate the importance of thoroughly testing any assumptions about user behaviour or knowledge that such techniques rest on.

\bibliographystyle{ACM-Reference-Format}
\bibliography{rws-first-look}


\begin{thebibliography}{13}


\ifx \showCODEN    \undefined \def \showCODEN     #1{\unskip}     \fi
\ifx \showDOI      \undefined \def \showDOI       #1{#1}\fi
\ifx \showISBNx    \undefined \def \showISBNx     #1{\unskip}     \fi
\ifx \showISBNxiii \undefined \def \showISBNxiii  #1{\unskip}     \fi
\ifx \showISSN     \undefined \def \showISSN      #1{\unskip}     \fi
\ifx \showLCCN     \undefined \def \showLCCN      #1{\unskip}     \fi
\ifx \shownote     \undefined \def \shownote      #1{#1}          \fi
\ifx \showarticletitle \undefined \def \showarticletitle #1{#1}   \fi
\ifx \showURL      \undefined \def \showURL       {\relax}        \fi
\providecommand\bibfield[2]{#2}
\providecommand\bibinfo[2]{#2}
\providecommand\natexlab[1]{#1}
\providecommand\showeprint[2][]{arXiv:#2}

\bibitem[Ayenson et~al\mbox{.}(2011)]%
        {ayenson2011flash}
\bibfield{author}{\bibinfo{person}{Mika~D Ayenson},
  \bibinfo{person}{Dietrich~James Wambach}, \bibinfo{person}{Ashkan Soltani},
  \bibinfo{person}{Nathan Good}, {and} \bibinfo{person}{Chris~Jay Hoofnagle}.}
  \bibinfo{year}{2011}\natexlab{}.
\newblock \showarticletitle{Flash cookies and privacy II: Now with HTML5 and
  ETag respawning}.
\newblock \bibinfo{journal}{\emph{Available at SSRN 1898390}}
  (\bibinfo{year}{2011}).
\newblock


\bibitem[Jackson et~al\mbox{.}(2006)]%
        {jackson2006protecting}
\bibfield{author}{\bibinfo{person}{Collin Jackson}, \bibinfo{person}{Andrew
  Bortz}, \bibinfo{person}{Dan Boneh}, {and} \bibinfo{person}{John~C
  Mitchell}.} \bibinfo{year}{2006}\natexlab{}.
\newblock \showarticletitle{Protecting browser state from web privacy attacks}.
  In \bibinfo{booktitle}{\emph{Proceedings of the 15th international conference
  on World Wide Web}}. \bibinfo{pages}{737--744}.
\newblock


\bibitem[Jueckstock et~al\mbox{.}(2022)]%
        {jueckstock2022measuring}
\bibfield{author}{\bibinfo{person}{Jordan Jueckstock}, \bibinfo{person}{Peter
  Snyder}, \bibinfo{person}{Shaown Sarker}, \bibinfo{person}{Alexandros
  Kapravelos}, {and} \bibinfo{person}{Benjamin Livshits}.}
  \bibinfo{year}{2022}\natexlab{}.
\newblock \showarticletitle{Measuring the Privacy vs. Compatibility Trade-off
  in Preventing Third-Party Stateful Tracking}. In
  \bibinfo{booktitle}{\emph{WWW}}. \bibinfo{pages}{710--720}.
\newblock


\bibitem[Klein and Pinkas(2019)]%
        {klein2019dns}
\bibfield{author}{\bibinfo{person}{Amit Klein} {and} \bibinfo{person}{Benny
  Pinkas}.} \bibinfo{year}{2019}\natexlab{}.
\newblock \showarticletitle{DNS Cache-Based User Tracking.}. In
  \bibinfo{booktitle}{\emph{NDSS}}.
\newblock


\bibitem[Li et~al\mbox{.}(2015)]%
        {li2015trackadvisor}
\bibfield{author}{\bibinfo{person}{Tai-Ching Li}, \bibinfo{person}{Huy Hang},
  \bibinfo{person}{Michalis Faloutsos}, {and} \bibinfo{person}{Petros
  Efstathopoulos}.} \bibinfo{year}{2015}\natexlab{}.
\newblock \showarticletitle{Trackadvisor: Taking back browsing privacy from
  third-party trackers}. In \bibinfo{booktitle}{\emph{International Conference
  on Passive and Active Network Measurement}}. Springer,
  \bibinfo{pages}{277--289}.
\newblock


\bibitem[Mayer and Mitchell(2012)]%
        {mayer2012third}
\bibfield{author}{\bibinfo{person}{Jonathan~R Mayer} {and}
  \bibinfo{person}{John~C Mitchell}.} \bibinfo{year}{2012}\natexlab{}.
\newblock \showarticletitle{Third-party web tracking: Policy and technology}.
  In \bibinfo{booktitle}{\emph{2012 IEEE symposium on security and privacy}}.
  IEEE, \bibinfo{pages}{413--427}.
\newblock


\bibitem[Pochat et~al\mbox{.}(2018)]%
        {pochat2018tranco}
\bibfield{author}{\bibinfo{person}{Victor~Le Pochat}, \bibinfo{person}{Tom
  Van~Goethem}, \bibinfo{person}{Samaneh Tajalizadehkhoob},
  \bibinfo{person}{Maciej Korczy{\'n}ski}, {and} \bibinfo{person}{Wouter
  Joosen}.} \bibinfo{year}{2018}\natexlab{}.
\newblock \showarticletitle{Tranco: A research-oriented top sites ranking
  hardened against manipulation}.
\newblock \bibinfo{journal}{\emph{arXiv preprint arXiv:1806.01156}}
  (\bibinfo{year}{2018}).
\newblock


\bibitem[Roesner et~al\mbox{.}(2012)]%
        {roesner2012detecting}
\bibfield{author}{\bibinfo{person}{Franziska Roesner},
  \bibinfo{person}{Tadayoshi Kohno}, {and} \bibinfo{person}{David Wetherall}.}
  \bibinfo{year}{2012}\natexlab{}.
\newblock \showarticletitle{Detecting and defending against third-party
  tracking on the web}. In \bibinfo{booktitle}{\emph{Presented as part of the
  9th {USENIX} Symposium on Networked Systems Design and Implementation ({NSDI}
  12)}}. \bibinfo{pages}{155--168}.
\newblock


\bibitem[Snyder et~al\mbox{.}(2020)]%
        {snyder2020filters}
\bibfield{author}{\bibinfo{person}{Peter Snyder}, \bibinfo{person}{Antoine
  Vastel}, {and} \bibinfo{person}{Ben Livshits}.}
  \bibinfo{year}{2020}\natexlab{}.
\newblock \showarticletitle{Who filters the filters: Understanding the growth,
  usefulness and efficiency of crowdsourced ad blocking}.
\newblock \bibinfo{journal}{\emph{Proceedings of the ACM on Measurement and
  Analysis of Computing Systems}} \bibinfo{volume}{4}, \bibinfo{number}{2}
  (\bibinfo{year}{2020}), \bibinfo{pages}{1--24}.
\newblock


\bibitem[Solomos et~al\mbox{.}(2021)]%
        {solomos2021tales}
\bibfield{author}{\bibinfo{person}{Konstantinos Solomos}, \bibinfo{person}{John
  Kristoff}, \bibinfo{person}{Chris Kanich}, {and} \bibinfo{person}{Jason
  Polakis}.} \bibinfo{year}{2021}\natexlab{}.
\newblock \showarticletitle{Tales of favicons and caches: Persistent tracking
  in modern browsers}. In \bibinfo{booktitle}{\emph{Network and Distributed
  System Security Symposium}}.
\newblock


\bibitem[Soltani et~al\mbox{.}(2010)]%
        {soltani2010flash}
\bibfield{author}{\bibinfo{person}{Ashkan Soltani}, \bibinfo{person}{Shannon
  Canty}, \bibinfo{person}{Quentin Mayo}, \bibinfo{person}{Lauren Thomas},
  {and} \bibinfo{person}{Chris~Jay Hoofnagle}.}
  \bibinfo{year}{2010}\natexlab{}.
\newblock \showarticletitle{Flash cookies and privacy}. In
  \bibinfo{booktitle}{\emph{2010 AAAI Spring Symposium Series}}.
\newblock


\bibitem[Syverson and Traudt(2018)]%
        {syverson2018hsts}
\bibfield{author}{\bibinfo{person}{Paul Syverson} {and}
  \bibinfo{person}{Matthew Traudt}.} \bibinfo{year}{2018}\natexlab{}.
\newblock \showarticletitle{$\{$HSTS$\}$ Supports Targeted Surveillance}. In
  \bibinfo{booktitle}{\emph{8th USENIX Workshop on Free and Open Communications
  on the Internet (FOCI 18)}}.
\newblock


\bibitem[Tiwari and Trachtenberg(2019)]%
        {tiwari2019alternative}
\bibfield{author}{\bibinfo{person}{Trishita Tiwari} {and} \bibinfo{person}{Ari
  Trachtenberg}.} \bibinfo{year}{2019}\natexlab{}.
\newblock \showarticletitle{Alternative (ab) uses for $\{$HTTP$\}$ Alternative
  Services}. In \bibinfo{booktitle}{\emph{13th USENIX Workshop on Offensive
  Technologies (WOOT 19)}}.
\newblock


\end{thebibliography}

\balance

\ifpdf
  \pdfinfo{
    /Title        (...)
    /Author       (...)
    /Subject      (...)
    /Keywords     (..., ..., ...)
    /CreationDate (D:20150827110616Z)
    /ModDate      (D:20150827110616Z)
    /Creator      (LaTeX)
    /Producer     (pdfTeX)
  }
  \ifdefined\pdftrailerid
    \pdftrailerid{}
    \pdfsuppressptexinfo=15
  \fi
\fi
\end{document}